# Expression of Interest for Evolution of the Mu2e Experiment[†]

F. Abusalma[23], D. Ambrose[23], A. Artikov[7], R. Bernstein[8], G.C. Blazey[27], C. Bloise[9], S. Boi[33], T. Bolton[14], J. Bono[8], R. Bonventre[16], D. Bowring[8], D. Brown[16], D. Brown[20], K. Byrum[1], M. Campbell[22], J.-F. Caron[12], F. Cervelli[30], D. Chokheli[7], K. Ciampa[23], R. Ciolini[30], R. Coleman[8], D. Cronin-Hennessy[23], R. Culbertson[8], M.A. Cummings[25], A. Daniel[12], Y. Davydov[7], S. Demers[35], D. Denisov[8], S. Denisov[13], S. Di Falco[30], E. Diociaiuti[9], R. Djilkibaev[24], S. Donati[30], R. Donghia[9], G. Drake[1], E.C. Dukes[33], B. Echenard[5], A. Edmonds[16], R. Ehrlich[33], V. Evdokimov[13], P. Fabbricatore[10], A. Ferrari[11], M. Frank[32], A. Gaponenko[8], C. Gatto[26], Z. Giorgio[17], S. Giovannella[9], V. Giusti[30], H. Glass[8], D. Glenzinski[8], L. Goodenough[1], C. Group[33], F. Happacher[9], L. Harkness-Brennan[19], D. Hedin[27], K. Heller[23], D. Hitlin[5], A. Hocker[8], R. Hooper[18], G. Horton-Smith[14], C. Hu[5], P.Q. Hung[33], E. Hungerford[12], M. Jenkins[32], M. Jones[31], M. Kargiantoulakis[8], K. S. Khaw[34], B. Kiburg[8], Y. Kolomensky[3,16], J. Kozminski[18], R. Kutschke[8], M. Lancaster[15], D. Lin[5], I. Logashenko[29], V. Lombardo[8], A. Luca[8], G. Lukicov[15], K. Lynch[6], M. Martini[21], A. Mazzacane[8], J. Miller[2], S. Miscetti[9], L. Morescalchi[30], J. Mott[2], S. E. Mueller[11], P. Murat[8], V. Nagaslaev[8], D. Neuffer[8], Y. Oksuzian[33], D. Pasciuto[30], E. Pedreschi[30], G. Pezzullo[35], A. Pla-Dalmau[8], B. Pollack[28], A. Popov[13], J. Popp[6], F. Porter[5], E. Prebys[4], V. Pronskikh[8], D. Pushka[8], J. Quirk[2], G. Rakness[8], R. Ray[8], M. Ricci[21], M. Röhrken[5], V. Rusu[8], A. Saputi[9], I. Sarra[21], M. Schmitt[28], F. Spinella[30], D. Stratakis[8], T. Strauss[8], R. Talaga[1], V. Tereshchenko[7], N. Tran[2], R. Tschirhart[8], Z. Usubov[7], M. Velasco[28], R. Wagner[1], Y. Wang[2], S. Werkema[8], J. Whitmore[8], P. Winter[1], L. Xia[1], L. Zhang[5], R.-Y. Zhu[5], V. Zutshi[27], R. Zwaska[8]

06 February 2018

## Abstract

We propose an evolution of the Mu2e experiment, called Mu2e-II, that would leverage advances in detector technology and utilize the increased proton intensity provided by the Fermilab PIP-II upgrade to improve the sensitivity for neutrinoless muon-to-electron conversion by one order of magnitude beyond the Mu2e experiment, providing the deepest probe of charged lepton flavor violation in the foreseeable future. Mu2e-II will use as much of the Mu2e infrastructure as possible, providing, where required, improvements to the Mu2e apparatus to accommodate the increased beam intensity and cope with the accompanying increase in backgrounds.

[†] Inquiries should be directed to Mu2e-II-contacts@fnal.gov

1. *Argonne National Laboratory, Lemont IL, USA*
2. *Boston University, Boston MA, USA*
3. *University of California, Berkeley CA, USA*
4. *University of California, Davis CA, USA*
5. *California Institute of Technology, Pasadena CA, USA*
6. *City University of New York, New York NY, USA*
7. *Joint Institute of Nuclear Research, Dubna, Russia*
8. *Fermi National Accelerator Laboratory, Batavia IL, USA*
9. *Laboratori Nazionali di Frascati of INFN, Frascati, Italy*
10. *Istituto Nazionali di Fisica Nucleare, Genova, Itlay*
11. *Helmholtz Zentrum Dresden-Rossendorf, Dresden, Germany*
12. *University of Houston, Houston TX, USA*
13. *Institute of High Energy Physics, Protvino, Russia*
14. *Kansas State University, Manhattan KS, USA*
15. *University College London, London, United Kingdom*
16. *Lawrence Berkeley National Laboratory, Berkeley CA, USA*
17. *Istituto Nazionale di Fisica Nucleare, Lecce, Italy*
18. *Lewis University, Romeoville IL, USA*
19. *University of Liverpool, Liverpool, United Kingdom*
20. *University of Louisville, Louisville KY, USA*
21. *University di Marconi, Rome, Italy*
22. *University of Michigan, Ann Arbor MI, USA*
23. *University of Minnesota, Minneapolis MN, USA*
24. *Institute for Nuclear Research, Moscow, Russia*
25. *Muons Inc, Batavia IL, USA*
26. *Istituto Nazionale di Fisica Nucleare, Napoli, Italy*
27. *Northern Illinois University, DeKalb IL, USA*
28. *Northwestern University, Evanston IL, USA*
29. *Novosibirsk State University/Budker Institute of Nuclear Physics, Novosibirsk, Russia*
30. *Istituto Nazionale di Fisica Nucleare, Pisa, Italy*
31. *Purdue University, West Lafayette IN, USA*
32. *South Alabama University, Mobile AL, USA*
33. *University of Virginia, Charlottesville VA, USA*
34. *University of Washington, Seattle WA, USA*
35. *Yale University, New Haven CT, USA*

## Introduction

The Mu2e experiment, now under construction at FNAL, will search for the rare muon-to-electron conversion process, $\mu^- + {}^{27}_{13}\mathrm{Al} \rightarrow e^- + {}^{27}_{13}\mathrm{Al}$. If this process were detected it would be the first experimental evidence of Charged Lepton Flavor Violation (CLFV) and a definitive signal of physics beyond the Standard Model. We are submitting this Expression of Interest as a preface to a detailed proposal to upgrade the Mu2e experiment, to improve the sensitivity by an order of magnitude. The new experiment is referred to here as Mu2e-II.

The muon to electron conversion rate is usually expressed in terms of the ratio of the conversion rate to the ordinary muon capture rate, $R_{\mu e} = \frac{\Gamma(\mu^- + N(A,Z) \rightarrow e^- + N(A,Z))}{\Gamma(\mu^- + N(A,Z) \rightarrow \nu_\mu + N(A,Z-1)^*)}$. Currently the best experimental limit is for the conversion rate in a gold stopping target, from the SINDRUM II experiment at PSI (2006), $R_{\mu e}(Au) < 7\times10^{-13}$ (90% c.l.). The goal of the Mu2e experiment is a single event sensitivity with an aluminum target of $R_{\mu e} = 2.5\times10^{-17}$, with a background of less than 0.5 events. If we assume a null signal for the purpose of comparison with previous experiments, this single event sensitivity corresponds to a limit of $R_{\mu e}(Al) < 7\times10^{-17}$ (90% c.l.), *i.e.*, a four order of magnitude improvement in sensitivity over previous measurements.

The Mu2e-II upgrade goal is to capitalize on the investment of Mu2e by incisively probing discoveries or extending the search for new physics by an order of magnitude. These goals require that Mu2e-II hold the total background level to less than one event with a single event sensitivity of $2.5\times10^{-18}$, a one order of magnitude improvement over Mu2e.

Mu2e-II will use as much of the existing Mu2e infrastructure and apparatus as possible. Some upgrades will, however, be required. We have therefore developed an R&D plan, which is described below. Preliminary studies have already been completed. A Snowmass white paper [1] studied the potential of 1000 and 3000 MeV high intensity proton beams for future experiments. These studies directly address the potential feasibility of Mu2e-II. Many issues of operating at higher beam flux were examined therein; these are highlighted in the discussion below. The background yields and hence the achievable sensitivity from these studies are summarized in Table 1, which assumes a proton beam kinetic energy of 1000 MeV. These backgrounds change by less than 10% for a proton beam energy of 3 GeV. We are currently refining these studies for the PIP-II 800 MeV proton beam. In the simulations used to make these background estimates, the Mu2e geometry was retained, while the tracker straw walls were thinned from 15 μm to 8 μm thickness and $BaF_2$ crystals replaced the CsI crystals in the baseline calorimeter. The extinction was assumed to be $1\text{x}10^{-12}$. It is apparent that a lesser extinction requirement of $1\text{x}10^{-11}$ would suffice, since it would increase the background by less than 0.02 events, dominated by an increase in the radiative pion background.

| Category | Source | Events (Al) | Events (Ti) |
|---|---|---|---|
| Intrinsic | $\mu$ decay in orbit | 0.26 | 1.19 |
| | Radiative $\mu$ capture | <0.01 | <0.01 |
| Late Arriving | Radiative $\pi$ capture | 0.04 | 0.05 |
| | Beam electrons | <0.01 | <0.01 |
| | $\mu$ decay in flight | <0.01 | <0.01 |
| | $\pi$ decay in flight | <0.01 | <0.01 |
| Miscellaneous | Anti-proton induced | -- | -- |
| | Cosmic ray induced | 0.16 | 0.16 |
| **Total Background:** | | **0.46** | **1.40** |

*Table 1: Estimated background yields for the Mu2e-II experiment assuming an aluminum (Al) or a titanium (Ti) stopping target. These studies were performed for a proton beam energy of 1 GeV. The total uncertainty is about 20%. Reproduced from arXiv:1307.1168. Note that, unlike in the case of aluminum, the titanium analysis has not yet been rigorously optimized.*

## Charged Lepton Flavor Violation

Two longstanding questions in particle physics demand attention: Why are there three fermion families and why does experimental evidence show, so far, that charged lepton flavor is conserved? Quarks mix in the Standard Model and we know now that neutrinos can oscillate from one lepton flavor to another. Neutral lepton flavor non-conservation (neutrino mixing) is direct evidence today of physics beyond the Standard Model. Clearly it would be momentous to observe charged lepton flavor non-conservation as well; to date no such violation has been observed. Indeed, the branching fractions predicted in the Standard Model are far below any conceivable experimental sensitivity. Observation of an experimentally accessible signal requires physics beyond the Standard Model. Many compelling new physics models predict a measurable muon to electron conversion signal in the Mu2e/Mu2e-II sensitivity range.

Because the properties of lepton flavor are so central to the Standard Model and so sensitive to new physics, there has been an experimental imperative since the beginnings of our field to search for and discover charged lepton flavor violation. Experiments using muons have focused on searches for the free muon decays, $\mu^+ \to e^+\gamma$, $\mu^+ \to e^+e^-e^+$, and the coherent muon to electron conversion in nuclei, $\mu^-\mathrm{N} \to e^-\mathrm{N}$. The current experimental limits [2], at 90% c.l., are $\mathrm{Br}(\mu^+ \to e^+\gamma) < 4.2\times10^{-13}$, $\mathrm{Br}(\mu^+ \to e^+e^-e^+) < 4.3\times10^{-12}$, and $R_{\mu e}(\mu^-\mathrm{Au} \to e^-\mathrm{Au}) < 7\times10^{-13}$. Significant experimental efforts are underway around the world to improve all these limits. Each of these processes have a complementary dependence on models and measurements of all of them can incisively explore the relevant new physics parameter space. For example, the muon to electron conversion process and to some extent $\mu^+ \to e^+e^-e^+$, are expected to be highly sensitive to contact term (point) interactions, while $\mu^+ \to e^+\gamma$ is mostly sensitive to new physics in loops.

As noted previously, the goal of Mu2e is to improve the sensitivity of $R_{\mu e}$ by four orders of magnitude, and the goal of Mu2e-II is to extend the sensitivity by another order of magnitude. Examples of Standard

Model extensions where Mu2e has discovery potential include [3], supersymmetry with or without R-parity conservation, models with multiple Higgs, Z' models, leptoquark models, and extra dimension models. The proposed sensitivity probes energy scales in the thousands of TeV in some scenarios; far beyond the energy reach of LHC experiments. It is conceivable that Mu2e or Mu2e-II could see a signal even in the absence of new signatures from the LHC.

The specific physics objectives of Mu2e-II depend on the results of the Mu2e experiment. There are several possible outcomes:

- If Mu2e sees no signal, the increased sensitivity of Mu2e-II will further constrain new physics parameters and improve the prospects for detection of CLFV.
- If Mu2e sees a signal with less than 5$\sigma$ significance, Mu2e-II will allow us to definitively establish that signal.
- If Mu2e sees a large signal (greater than 5$\sigma$), Mu2e-II could make a precision measurement of the effect and operate with different stopping targets (such as Ti). Measuring the *Z* dependence of $R_{\mu e}$ is a unique window afforded by the muon-to-electron conversion process and can probe the structure of new physics as shown in Figure 1.

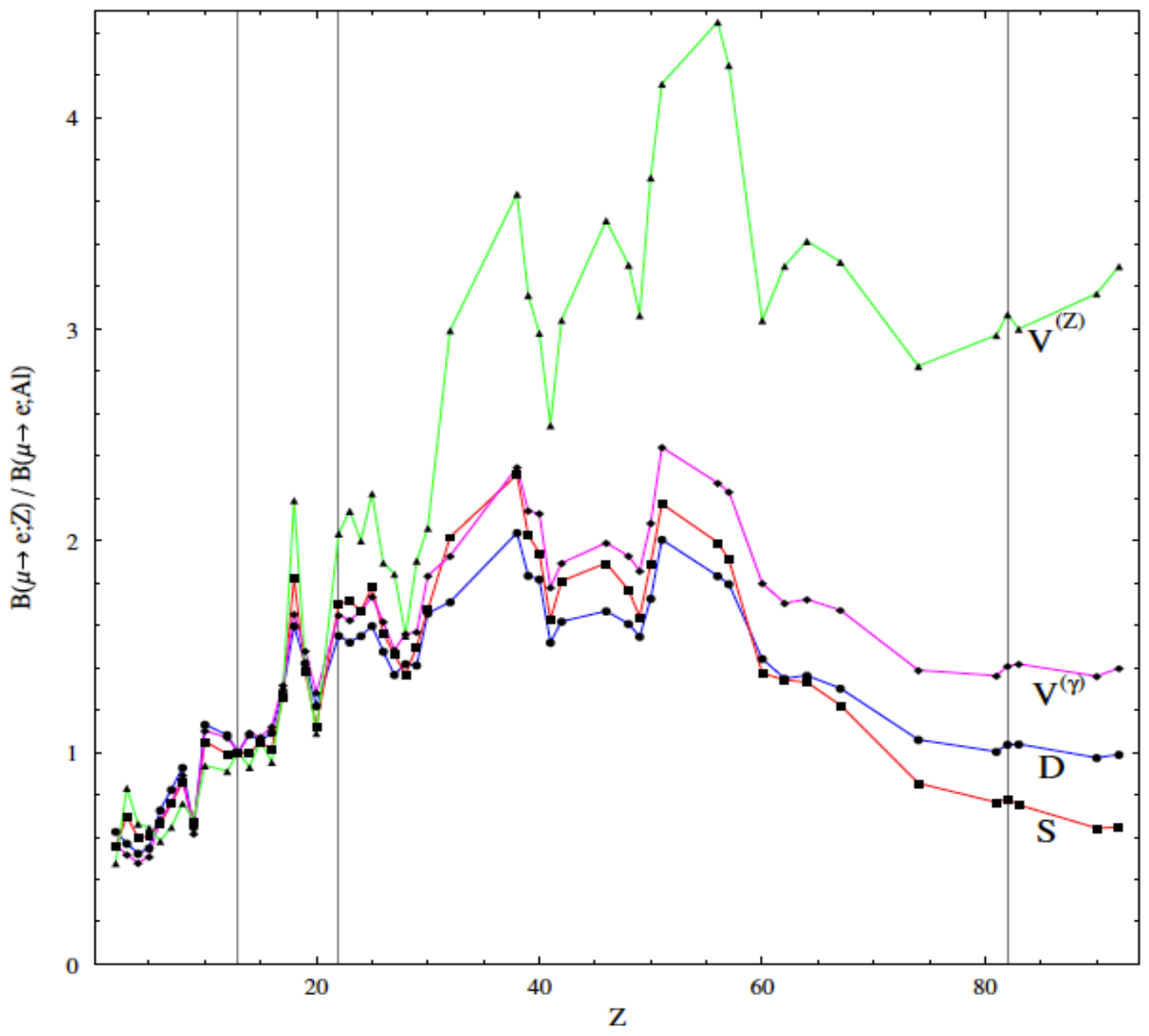

*Figure 1. The variation of $R_{\mu e}$ as a function of the Z of the stopping target depends on the dominant operator in the Lagrangian. Measuring $R_{\mu e}$ of different atomic stopping targets can help distinguish among new physics operators (S, D, $V^{(\gamma)}$, $V^{(Z)}$) [4].*

## Additional New Physics Searches

In addition to the search for lepton flavor violation, Mu2e-II will offer unique sensitivity to a lepton *number* violating muon-to-positron conversion $\mu^{-}+A(Z, N) \rightarrow e^{+}+A(Z-2, N+2)$. This process is complementary to searches for neutrinoless double beta decay (0νββ), although it is sensitive to specific models of new physics which may not manifest in 0νββ. Both processes can proceed through the exchange of a virtual, massive Majorana neutrino [5]. Unlike 0νββ, which involves coupling of the massive neutrinos to the electron, the muon-to-positron conversion is sensitive to the lepton couplings between the second and first generations, and depends on a different linear combination of the neutrino masses. Thus, the two processes measure a complementary set of parameters. Observation of both processes would provide information on the otherwise inaccessible Majorana phases in the neutrino mixing matrix. Should the lepton number be violated by a new physics mechanism other than massive Majorana neutrinos, the two processes would sample a different set of new physics couplings.

The current best limit on negative muon-to-positron conversion comes from the SINDRUM experiment at PSI [6]:

$$R_{\mu e+} = \frac{\Gamma(\mu^- + ^{48}\mathrm{Ti} \rightarrow e^+ + ^{48}\mathrm{Ca}^{GS})}{\Gamma(\mu^- + ^{48}\mathrm{Ti} \rightarrow \nu_\mu + ^{48}\mathrm{Sc}^*)} < 1.7*10^{-12}$$

Mu2e-II sensitivity to $R_{\mu e+}$ will depend on the choice of the target material. The ideal target nucleus would have the largest mass difference between the initial *A(*Z*,* N*)* and final *A(*Z*−2,* N*+2*) nuclear states, and therefore the largest possible $e^+$ energy, compared to the dominant background from the radiative muon capture process. While $^{27}$Al is one of the better candidates for the $\mu \rightarrow e^-$ conversion, $^{28}$Si, $^{48}$Ti, and $^{40}$Ca have the best sensitivity to $\mu \rightarrow e^+$. Active silicon target technology, likely requiring a dedicated lower stopping rate special run, may be particularly appealing for Mu2e-II. Initial estimates indicate that Mu2e-II would improve the sensitivity to lepton number violating $\mu^- \rightarrow e^+$ conversion by 2-3 orders of magnitude compared to the current limits.

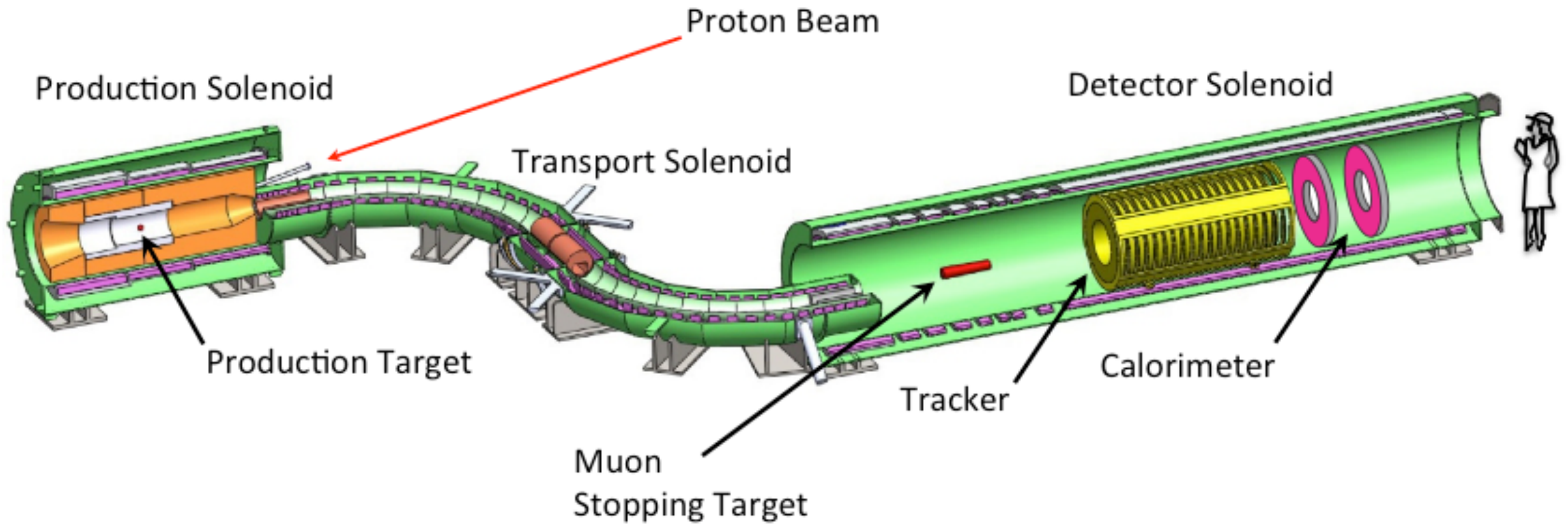

*Figure 2. The Mu2e experimental layout.*

## The Experimental Strategy of Mu2e and Mu2e-II

The muon beamline consists of three solenoids: the Production Solenoid (PS), the Transport Solenoid (TS), and the Detector Solenoid (DS), connected in series (see Figure 2). The magnets form a continuous magnetic field that decreases with varying gradients from the muon production target to the muon stopping target. The negative magnetic gradient guides particles downstream, improves the collection of particles by mirroring some upstream bound particles back downstream, and, as explained below, prevents the loitering of particles along the beam line that could create background in the delayed measurement time window.

Protons strike a production target in the middle of the Production Solenoid, producing low momentum pions that decay to muons. Captured muons spiral along the magnetic field lines from the PS through the TS, eventually stopping in a thin target in the DS. The S-shaped TS serves to eliminate a straight-line path for neutral particles from the Production Target to the detectors, selects the desired particle charge, and narrows the momentum range of captured muons. Charge selection is accomplished through toroidal transport, in which captured particles drift vertically up or down, depending on the sign of the charge with a displacement that depends on momentum. A portion of the muons stop in the thin stopping targets (aluminum) located near the upstream end of the DS while the balance of the beam is transported to a downstream beam dump. The stopped muons are captured immediately into aluminum atomic orbits, quickly settling in the K-shell to form a muonic atom, which has a lifetime [7] of 864 ns - a relatively long time - during which the muon has a large overlap with the nucleus that can host a new physics interaction. Note that an interacting nucleus is required to conserve energy and momentum in the coherent muon-to-electron conversion process. The muon beam line for Mu2e delivers about 0.002 stopped muons per incident proton, making it much more efficient for collecting muons than conventional beam lines consisting of quadrupoles and bending magnets.

The three main interactions of the muon in a muonic aluminum atom (the chosen target material for Mu2e) are:

1) The conversion process, $\mu^- + {}^{27}_{13}\mathrm{Al} \rightarrow e^- + {}^{27}_{13}\mathrm{Al}$, resulting in a conversion electron energy of 104.96 MeV.

2) Muon decay in orbit (DIO), $\mu^- + {}^{27}_{13}\mathrm{Al} \rightarrow e^- + \nu_\mu + \bar{\nu}_e + {}^{27}_{13}\mathrm{Al}$, with a 39% branching ratio. If the muon were free, the maximum electron energy would be 53 MeV. However, due the presence of a nearby nucleus that can absorb momentum and energy, it is possible for the neutrinos to have zero energy and consequently the electron energy can approach that of a conversion electron. This process poses a significant background to the conversion electron signal (see Figure 3). Fortunately, the $e^-$ rate falls rapidly with increasing energy near the endpoint energy, and the background can therefore be controlled with good electron energy resolution. Most of these electrons have energies less than 53 MeV, and spiral harmlessly along the solenoid axis through the holes in the Tracker and the Calorimeter to a beam dump, thereby minimizing the background rate in these detectors.

3) The muon captures on the nucleus, $\mu^- + {}^{27}_{13}\mathrm{Al} \rightarrow \nu_\mu + {}^{27}_{12}\mathrm{Mg}^*$, via the weak interaction, with a 61% branching ratio. The excited magnesium nucleus can decay via several channels, including the prompt emission of photons, neutrons, protons, deuterons, *etc.*, which leads to backgrounds and possibly radiation damage in the detectors.

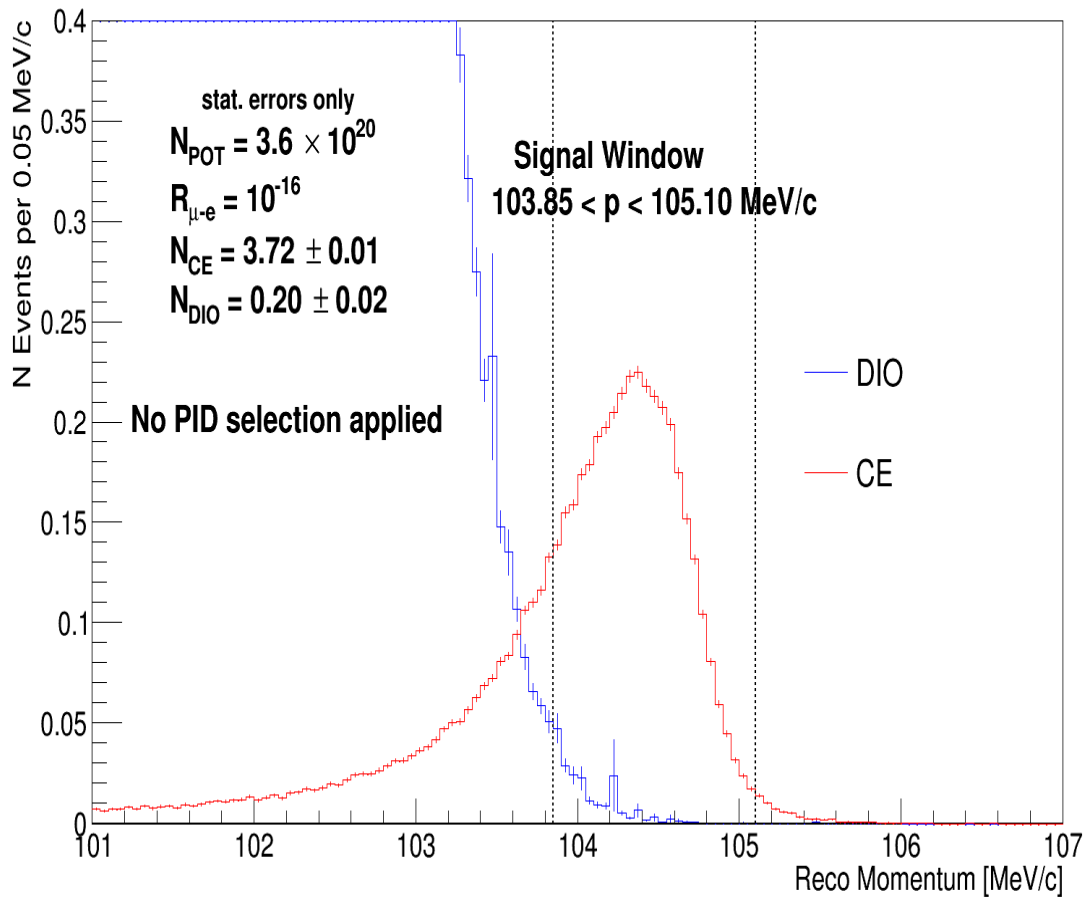

*Figure 3. A simulation of the electron spectra from the straw tube tracker. The conversion electron signal (in red) assumes* $R_{\mu e} = 1 \times 10^{-16}$. *The electron background from muons decaying in orbit (DIO) is in blue.*

The Snowmass study also investigated a next generation Mu2e experiment based on other stopping targets such as titanium. The study concluded that operation with a titanium target is plausible and would be an important tool (as evident in Figure 1) to understand the nature of any signal for new physics observed in Mu2e. The estimated background yields for a titanium stopping target are given in Table 1. The study used the geometry and algorithms optimized for the aluminum target except that the total mass of the titanium stopping target was made equal to the mass of the current default aluminum stopping target by varying the thickness and number of stopping target foils. A rigorous optimization for a titanium stopping target has not yet been performed, but might yield a reduced number of DIO background events while maintaining a similar signal acceptance.

## Proton Beam Requirements

The Mu2e-II experiment requires a high frequency (500-2000 kHz) pulsed proton beam with ten to twenty times more integrated power than Mu2e. The beam energy must be high enough to efficiently produce charged pions that subsequently decay to muons, and preferably safely below the anti-nucleon production threshold (6-8 GeV kinetic energy) to avoid an important class of backgrounds. The Mu2e proton beam kinetic energy is constrained to 8 GeV by the MI-8 beamline and Recycler Ring permanent magnets. The Snowmass study and subsequent work [8] has demonstrated that beam kinetic energies between 800 MeV and 4000 MeV optimize the single event sensitivity for the Mu2e-II experiment as seen in Figure 4.

A pulsed beam is required to eliminate a major background from pions halting and decaying in the muon stopping target. The pions are unavoidably contained in the muon beam line. The pions and muons arrive at the stopping target shortly after the proton pulse strikes the production target. In both Mu2e and Mu2e-II, the data measurement window is delayed for about 700 ns after the proton pulse to provide time for the vast majority of the pions to decay or annihilate in material. However, because of their relatively long lifetime, most of the muonic atoms remain.

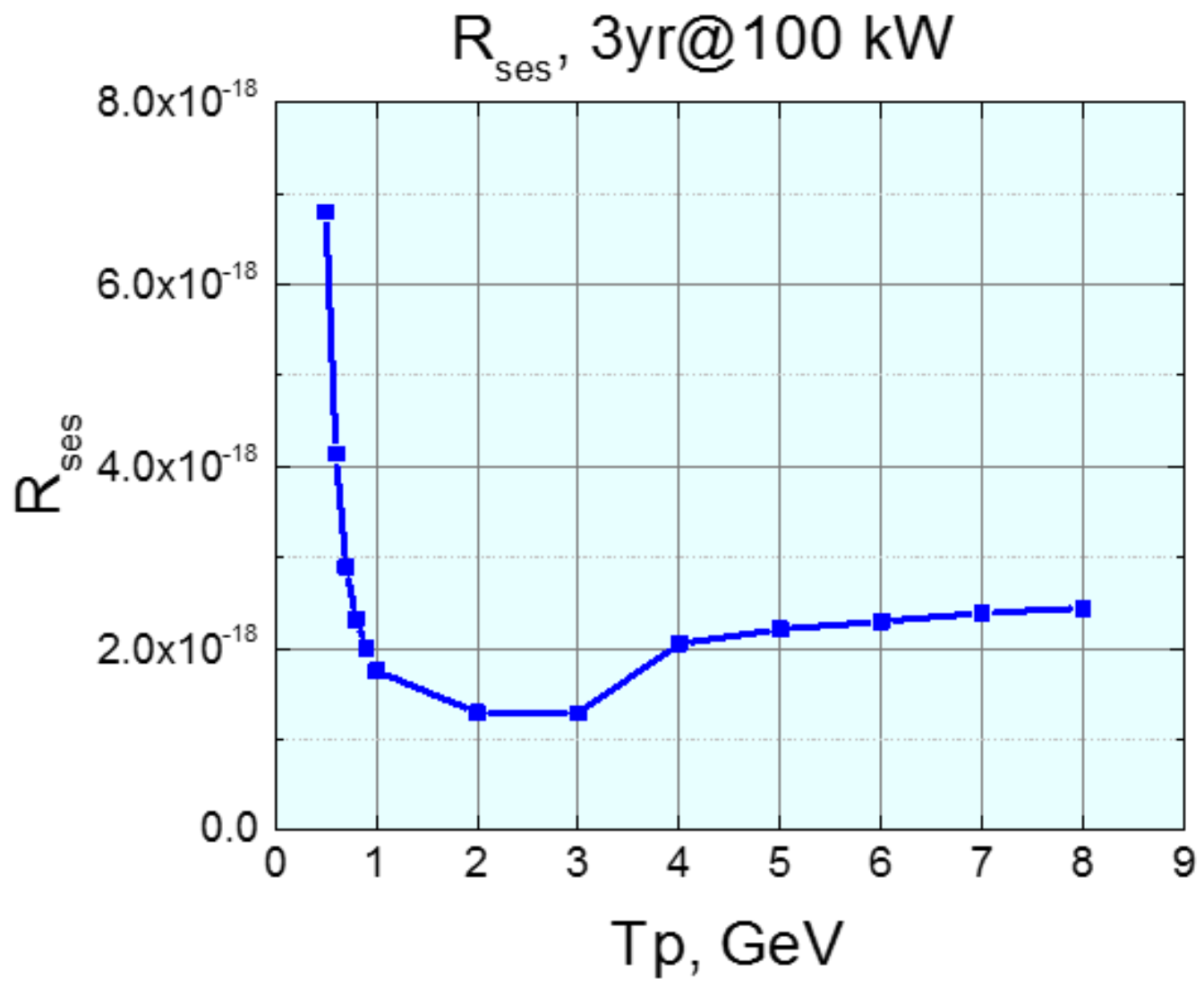

*Figure 4: An estimate of the single-event-sensitivity (SES) for Mu2e-II for a nominal 3-year run at 100 kW for various proton beam energies (Tp). The nominal Mu2e SES is 2.5 x $10^{-17}$ for 3 years of 8 kW proton beam at 8 GeV. This curve assumes that the limitations discussed in [8] can be mitigated, that the detector performance remains unchanged at the higher beam intensities, and that an aluminum stopping target is used. Taken from arXiv:1612.08931.*

The proton pulses must be narrow, ideally < 100 ns base width (the Mu2e pulse width is 250 ns, which is wider than optimal), and ideally separated by a time which can be varied from 800 to 2000 ns. The microstructure of the beam (structure inside the < 100 ns-wide pulse) is not important.

There must be almost no proton beam between pulses to avoid the production of background-producing pions during the measurement window. Mu2e-II requires that the integrated number of protons between pulses be a factor of $1\times10^{11}$ less than the number in the pulse (called the extinction factor). Mu2e will achieve $1\times10^{10}$ suppression by having at least a factor of $10^5$ from extinction in the Recycler Ring and Delivery Ring, and another factor of $10^6$ -$10^7$ provided by an AC dipole magnet-collimator placed just upstream of the Mu2e proton target. Simulations suggest [11] this joint approach can be further optimized to achieve an extinction factor of $1\times10^{11}$ for Mu2e-II with the intrinsic PIP-II linac extinction and the AC dipole extinction.

The fraction of the high frequency pulse train in the time line (macro duty-factor) should be as high as possible; optimally greater than 90%. Additionally, the pulse train should have minimal pulse-to-pulse variation throughout the train, optimally less than 10%. A new beamline will be required to transport proton beam from PIP-II to the Mu2e production target station. In the Mu2e scheme, a proton bunch is stored in the Delivery Ring (evolved from the Fermilab anti-proton complex), and beam is resonantly extracted on every turn and transported to the Mu2e target. The higher beam power required by Mu2e-II cannot be realized without the addition of substantial shielding above the Delivery Ring, dictated by radiation levels in human habitation zones nearby. In addition, it will be extremely difficult for the Mu2e slow-extraction scheme to achieve narrower pulses, improve extinction, and increase pulse-to-pulse uniformity at the beam power level required for Mu2e-II.

Mu2e will operate with an 8 kW beam at 8 GeV proton kinetic energy, with 250 ns wide (base width) proton pulses separated by 1695 ns. Within the accelerator complex meta-cycle, pulses are delivered to Mu2e for 0.38 seconds followed by a beam-off period of 1.02 seconds, for an effective 25% macro duty factor. The total number of muons stopping within the Mu2e detector will be of order $10^{18}$ in three years of running. Mu2e-II, using protons from PIP-II at 800 MeV, will require about 100 kW to deliver $10^{19}$ stopped muons in three years of data taking. Realizing an improved macro duty factor goal of higher than 90% will limit the increase in instantaneous detector rates for Mu2e-II to about three times the nominal Mu2e instantaneous rates.

## Muon Beam Line

The radiatively-cooled tungsten target [12] used for Mu2e must be replaced to handle the higher power deposited in the target. Options that have been studied [13] for handling the higher beam power while still maintaining a sufficient stopped-muon yield include employing active cooling (e.g. water or helium gas), using a liquid target, and/or rasterizing the beam on the target face.

For Mu2e-II, the Production Solenoid will require an improved heat and radiation shield between the production target and the cryostat, to maintain acceptable levels of radiation damage and heat load in the superconducting coils. Simulations demonstrate that changing the Mu2e shield from brass to tungsten will provide adequate thermal protection for the superconducting coils, but the DPA (displacements per atom, a measure of radiation damage) is still uncomfortably large, by a factor of 2 or 3. Although the DPA damage can be annealed at room temperature, we want to avoid annealing more than about once per year. However, preliminary studies suggest that upgrading to a tungsten shield may not be sufficient to avoid more frequent annealing. Replacement of the entire production solenoid and shielding may be required for Mu2e-II. Further study of radiation damage to the PS at high beam power is required. Based on our current understanding of radiation damage and heat load in the superconducting coils, no upgrades of either the Transport or Detector Solenoids appear to be necessary.

## Detector Considerations for Mu2e-II

### Tracker

The Tracker provides the primary information on the momentum of putative conversion electrons, as they spiral through the DS magnetic field. The Tracker must provide excellent momentum resolution in order to separate the conversion electrons from the high-energy tail of the DIO electrons (see Figure 3). For Mu2e, this background is a manageable 0.2 events over the life of the experiment, with a total momentum resolution, including energy straggling effects in upstream material, of about 400 keV/c for 105 MeV/c electrons originating from the stopping target. With the same Tracker in Mu2e-II, however, this background exceeds two events, becoming the dominant background and exceeding the proposed total error budget of 1 event at the Mu2e-II sensitivity goal. The Snowmass study indicated that tracker performance would be robust against a factor of two or three increase in the instantaneous rate. However, achieving the background goals for Mu2e-II will require an improved spectrometer resolution. In Mu2e, for a 105 MeV/c electron, the dominant contributions to the spectrometer resolution come from energy loss straggling in the stopping target, the inner proton absorber, and straggling and scattering in the tracker material itself. The low-mass tracker [14] for Mu2e is composed of straw planes with 15-micron straw wall thickness. Preliminary studies indicate that the spectrometer resolution required to keep the Mu2e-II DIO background under control can be achieved by utilizing straws with a wall thickness of 8 microns. The x10 integrated radiation dose on the Mu2e-II tracker electronics motivates study of appropriate rad-hard readout electronics at a level informed by the HL-LHC detector upgrades and beyond. Whereas HL-LHC detectors are primarily concerned with Single Event Upsets (SEUs), Mu2e-II readout electronics will be challenged by both SEU and total dose sensitivity.

## Cosmic Ray Veto (CRV)

The Cosmic Ray Veto (CRV) is a plastic scintillator system [14] that hermetically surrounds the Detector Solenoid to veto cosmic ray muons that can induce background signals. This cosmic-induced background scales with live-time which will be three to four times greater in Mu2e-II compared to Mu2e for the same nominal 3 year run. Noise hits in the CRV can lead to false vetoes, reducing the efficiency for the detection of conversion electrons. A major source of noise hits is neutrons produced by nuclear muon capture in the stopping target, collimators, and muon beam stop. It is likely that the shielding between these sources and the CRV will have to be improved for Mu2e-II to have an acceptable accidental veto rate of less than 10%. The x10 integrated radiation dose on the Mu2e-II CRV readout electronics motivates study of appropriate rad-hard readout electronics at a level informed by the HL-LHC detector upgrades.

## Calorimeter

The Mu2e calorimeter [14] consists of pure CsI crystals comprising two disks as indicated in Figure 2. The calorimeter has robust rate performance at Mu2e rates but may be challenged by Mu2e-II instantaneous rates that are two to three times higher. The x10 integrated radiation dose on the calorimeter readout electronics motivates study of appropriate rad-hard readout electronics at a level informed by the HL-LHC detector upgrades. An alternative calorimeter design has been developed based on $BaF_2$ crystals readout with solar-blind UV sensitive avalanche photo-diodes or SiPMs that efficiently collect the very fast UV component (~220 nm) of the scintillation light while suppressing the slow component near 310 nm. This alternative design would be considerably more robust against Mu2e-II rates but requires the development and commercialization of the required solid state photo sensors, which is ongoing.

## DAQ

The data acquisition system will need to be upgraded to handle the 10x larger throughput of Mu2e-II. Handling the higher data volume and storage requirements should be possible with anticipated improvements in technology. In addition, the software-based trigger filter will require improved performance to handle the higher instantaneous rates and larger duty factor. Further investigation of these issues is required.

## PIP-II Project

We assume that Mu2e-II will be developed on the same time scale as the PIP-II Project, with projected beam operations in the late 2020's. The PIP-II Project [11] has recently received "Critical Decision 1" (CD1) recognition from the DOE Office of Science. The CD1 process included external review of four different linac technologies which were down-selected to a baseline design of a high duty factor superconducting RF linac capable of providing the proton beam power, extinction, and pulse-train required by Mu2e-II.

## Mu2e-II R&D plan

Progressing the Mu2e-II Expression of Interest to a proposal requires advancing the plausibility arguments and studies presented through an R&D plan that addresses the leading technical issues and risks. The R&D plan can advance on both the "Grand Challenges" identified by the detector R&D community [9] and the targetry R&D goals identified by the General Accelerator R&D (GARD) HEPAP sub-panel [10]. The leading R&D issues and associated plans are summarized below.

### Required extinction R&D and beam transport simulations

The proton beam for both Mu2e and Mu2e-II is injected off-axis into the Production Solenoid as shown in Figure 2. The Mu2e-II beam kinetic energy (800 MeV) is an order of magnitude less than the Mu2e beam energy (8000 MeV) and will hence traverse a much different trajectory through the Production Solenoid field. Preliminary studies have demonstrated that Mu2e-II off-axis beam injection and transport may be possible, but considerable work remains to demonstrate that off-axis injection at this much lower energy is credible.

High-field linear accelerating structures are capable of intrinsically high extinction and levels approaching $10^9$ have been demonstrated in simulation [11]. Demonstrating a high level of extinction with the "PIP2IT" R&D platform for PIP-II is an important goal, as is developing a strategy for achieving the required total joint extinction of $10^{-11}$ for Mu2e-II with the nominal AC-dipole/collimator system. The PIP-II linear accelerator nominally accelerates $H^-$ ions to facilitate injection into synchrotrons of the future accelerator complex. R&D is required to demonstrate that $H^-$ ions can be efficiently transported to the Mu2e-II target. If $H^-$ stripping to $H^+$ is needed then R&D on low-loss stripping techniques will likely be required for the Mu2e-II transfer line.

### Required proton target R&D

The Mu2e-II proton target environment will be much more severe than that of Mu2e. To achieve the roughly 10× increase of beam power using a lower energy proton beam will require an approximately 120× increase in the number of incident protons per year. This is a new regime in high power targetry with an irradiation density exceeding any target at any facility to date. The higher duty factor of Mu2e-II will also complicate beam-abort schemes. Pre-conceptual designs of helium and water cooled target systems exist [13], and will need to be advanced to Mu2e-II conditions. Collaboration with the RaDIATE high target power R&D consortium pursuing GARD targetry R&D goals will be invaluable. The existing beam dump and associated cooling systems also require study.

## Required study of radiological issues

Two principle sources of radiological hazard during the operation of the Mu2e experiment have been identified: (1) radiation from beam loss during proton resonant extraction from the Delivery Ring, and (2) prompt radiation from the interaction of the proton beam with the production target and absorber. Radiation from these sources will reach accessible areas either directly or from the sky-shine that results from interaction of secondary neutrons in the atmosphere producing radiation far from the source.

Radiation dose rates for 8 kW Mu2e operation have been estimated to be 3-5 mRem/hr direct dose on the berm downstream of the proton target and less than 0.1 mRem/year dose at Wilson Hall due to sky-shine from the Mu2e proton target [15].

Mu2e-II will not use the Delivery Ring, thus eliminating the leading radiological hazard. However, the increased beam power on the proton target will increase both the direct and sky-shine dose rates from the Mu2e building. Moreover, the increased beam power at lower energy and at a higher duty factor entails about a 120× increase in the average number of protons per second incident on the target. Consequently, target activation and the beam-off radiation levels will be larger than that anticipated for 8 kW Mu2e operation.

The upgrade path to Mu2e-II will likely require the replacement of the Production Solenoid/Heat and Radiation Shield assembly. Several years of running the Mu2e experiment will make this a “hot job” – work on highly radioactive material in a room with large residual radioactivity. Calculations are needed to estimate the cooldown time required to reduce the activation of the Production Solenoid interior and the residual radioactivity of the Mu2e target hall to acceptable levels.

R&D is required to simulate the radiological environment of Mu2e-II and to determine the upgrades necessary to maintain compliance with the Fermilab Radiological Controls Manual.

## Required dose/DPA studies for the Mu2e-II PS

The muon production yield and associated radiation damage to the Mu2e Production Solenoid has been studied in some depth [8]. Further work is motivated to study hybrid Heat and Radiation Shield (HRS) designs, such as layered High-Z/CH structures for example, at Mu2e-II beam energies. The Mu2e-II HRS radiation dose profile with 800 MeV beam will be much more central than the Mu2e profile with 8 GeV beam. Hence the design of the HRS, dump and location of the Mu2e-II target within the Production Solenoid will need to be carefully optimized for Mu2e-II.

The excitation field of the Production Solenoid will be limited by Residual Resistivity Ratio (RRR) in the pure aluminum quench matrix of the superconducting cable. Radiation from the production target damages the quench matrix and this damage is codified as “Displacements Per Atom” or DPA. Calibrating the relationship between beam power on target with DPA and the limiting RRR factor is a continuing R&D program that will be informed by both early running of the COMET Stage-1 program at JPARC [16] and early running of Mu2e.

## Required R&D for the cosmic veto system

The Mu2e Cosmic Ray Veto (CRV) and calorimeter readout system are based on Geiger-mode avalanche photodiodes referred to as "Silicon Photo-Multipliers" or "SiPMs". SiPMs are well suited to operating in a high magnetic field, but are not particularly radiation hard, resulting in substantial yet tolerable dark currents in Mu2e. For Mu2e-II the nominal Mu2e CRV shielding will not suffice for Mu2e operations with a x10 integrated dose. R&D on the benefits of additional shielding and cooling [17] of the SiPM readout system to reduce dark currents is required in order preserve veto efficiency. The radiation damage from the x10 integrated dose on CRV scintillator and fibers likewise needs to be understood; R&D may be required to mitigate aging effects.

## Required R&D for calorimetry

The Mu2e-II environment presents two challenges to the calorimeter system. First, the x10 increase in integrated dose (principally neutrons) corresponds to 10 kGy ( $1x10^{13}$ $n/cm^2/sec$) for both crystals and sensors motivates consideration of more radiation tolerant crystals and sensors such as $BaF_2$ and Solar-blind Avalanche Photo-Diodes (APDs) and SiPMs. Second, the x3 increase in instantaneous rate motivates faster readout schemes such as utilizing the very fast component (0.9 nsec) of the $BaF_2$ UV scintillator light while suppressing or rejecting the larger, relatively slow (600 nsec) longer wavelength scintillation component with so called "solar-blind" filters.

Promising filter R&D has begun [18] at JPL/Caltech on integrated interference filters on high-speed rad-hard APDs and SiPMs. More speculative R&D on Gallium Nitride micro-channel plate readout technology [20] is also being studied to understand the speed and radiation hardness of this intriguing technology.

R&D has also been carried out to suppress the $BaF_2$ slow scintillation component by introducing rare earth doping in crystals [19]. Recent progress in yttrium doped $BaF_2$ shows a significant increase in the ratio between the fast and slow scintillation components from 1/5 to 5/1, while maintaining the amount of the fast light in $BaF_2$, which is similar to the yield of un-doped CsI [20]. This calorimetry R&D portfolio advances both the "Large Area Photodetectors" and the "Picosecond time barrier" Grand Challenges identified by the detector R&D community [9].

## Required R&D for tracker

Lower mass straws required for Mu2e-II will require another round of aging, sag and leak studies. The Mu2e tracker readout electronics is based on commercial components that are not particularly robust against integrated dose radiation damage, and are estimated to become inoperable at the x10 total dose currently estimated for Mu2e-II. The Mu2e collaboration has studied a radiation hard ASIC front-end solution for the tracker readout that could benefit Mu2e-II. Radiation-hard data transmission optical links will also be required for Mu2e-II. This requires radiation hardness R&D from similar ongoing studies for the High Luminosity LHC detector upgrades. R&D is also required to explore whether the Mu2e straws can tolerate a x10 increase in total charge, and, if not, how to mitigate aging effects. We note that thinner walls will reduce the charge load on the straws, as the dominant source of ionization is photons converting in the straw walls. Exploring other tracking technologies may be motivated if insurmountable issues are encountered with lower mass straws and readout in higher radiation fields, and this exploration could advance the “Ultra-low mass/power rad-hard silicon detectors” Grand Challenge identified by the detector R&D community [9].

## Opportunities to collaborate with COMET Stage-1

COMET is a competing muon-to-electron conversion research program being developed in Japan, that will follow a staged approach. The first stage has a sensitivity goal x100 better than present limits [2]. The second stage would compete at the Mu2e sensitivity level. The current COMET Stage-1 plan [16] calls for commissioning running in 2019 and first physics running in 2020 with 3.2 kW of 8000 MeV proton beam power. There may be an opportunity for joint R&D with Mu2e and Mu2e-II on Dose-DPA-RRR measurements and cross-calibrations.

## Summary

The Mu2e-II initiative is well matched to the opportunities presented by the Fermilab PIP-II accelerator upgrade, community pursuit of detector R&D “Grand Challenges”, investments in experimental infrastructure from the Mu2e Project and the growing strength of the experimental muon physics community. Mu2e-II can either incisively probe discoveries made by Mu2e or search more deeply for charged lepton flavor violation. In either case this initiative is uniquely positioned to pursue the mystery of charged lepton flavor conservation in the coming decades.

# References

[1] "Feasibility Study for a Next Generation Mu2e Experiment", K. Knoepfel, el al., arXiv:1307.1168, October 1, 2013. https://arxiv.org/abs/1307.1168 .

[2] A. Baldini, et al. [MEG Collaboration], Eur. Phys. J. C76: 434 (2016). Particle Data Group for the other limits, http://pdg.lbl.gov/ .

[3] Y. Kuno, Y. Okada, Rev. Mod. Phys. 73, 151 (2001); W. Marciano, T. Mori, Roney, Ann. Rev. Nucl. Sci. 58 (2008); M. Raidal et al, Eur. Phys. J. C57: 182, (2008); A. de Gouvea, P. Vogel, arXiv:1303.4097 [hep-ph] (2013).

[4] V. Cirigliano, *et al*., PRD 80, 013002 (2009).

[5] S. R. Elliott and P. Vogel, Ann. Rev. Nucl. Part. Sci. 52, 115 (2002).

[6] J. Kaulard *et a*l. [SINDRUM II Collaboration], Phys. Lett. B 422, 334 (1998).

[7] Measurements of Muon Disappearance Rates vs Time in C, Mg, Al, Si, and P J. L. Lathrop, R. A. Lundy, V. L. Telegdi, R. Winston, and D. D. Yovanovitch Phys. Rev. Lett. 7, 107 (1961)

[8] "A Study Of The Energy Dependence Of Radiation Damage In Superconducting Coils For a Next Generation Mu2e At PIP-II". V. Pronskikh *et al*., https://arxiv.org/abs/1612.08931 .

[9] Coordinating Panel for Advanced Detectors (CPAD): https://www.anl.gov/hep/initiatives/coordinating-panel-advanced-detectors , see also the recent CPAD presentation to HEPAP where the "Grand Challenges" are discussed: https://science.energy.gov/~/media/hep/hepap/pdf/201703/Demarteau_cpad_report_hepap.pdf

[10] General Accelerator R&D HEPAP sub-panel report: https://science.energy.gov/~/media/hep/hepap/pdf/Reports/Accelerator_RD_Subpanel_Report.pdf

[11] PIP-II Conceptual Design Report, http://pip2-docdb.fnal.gov/cgi-bin/ShowDocument?docid=113

[12] The Mu2e radiatively cooled target design is described in the Mu2e TDR. Presentations on the Mu2e target design can be found at the recent RAL/Oxford high power target workshop: https://eventbooking.stfc.ac.uk/news-events/6th-high-power-targetry-workshop-309?agenda=1 .

[13] Studies of Helium cooled target designs: Tristan Davenne, Mu2e-doc- 4064, 5520, 6790.

[14] Mu2e Technical Design Report, https://arxiv.org/abs/1501.05241 .

[15] A. Leveling, "Final design review for Accelerator Radiation Safety Improvements" Mu2e-doc-7145

[16] COMET Conceptual Design Report: https://inspirehep.net/record/842353/files/comet-cdr-v1.0.pdf COMET Phase-I Technical Design Report: http://comet.kek.jp/Documents_files/IPNS-Review-2014.pdf

[17] "Scintillating Fibre Tracking at High Luminosity Colliders", C.Joram *et al.*,2015, JINST 10, C08005.

[18] "An APD for the efficient detection of the fast scintillation component of BaF2", D.Hitlin *et al.*, NIM-A, Vol 824, p 119-122 (2016).

[19] "Development of $BaF_2$ Crystals for Future HEP Experiments at the Intensity Frontiers", F. Yang, J. Chen, L. Zhang and R.Y. Zhu, Paper N36-7 in 2016 IEEE NSS/MIC Conference Record. http://www.hep.caltech.edu/~zhu/papers/16_N36-7.pdf

[20] "Applications of Very Fast Inorganic Crystal Scintillators in Future HEP Experiments", R.-Y. Zhu, Talk given in the International Conference on Technology and Instrumentation in Particle Physics 2017 (TIPP2017), Beijing, China, to be published in Springer Proceedings in Physics. http://indico.ihep.ac.cn/event/6387/session/11/contribution/175/material/slides/0.pdf